\documentclass[prb,preprint]{revtex4}
\usepackage{graphicx}
\begin{document}
\title{Quantum corrections to the spin-wave spectrum of La$_2$CuO$_4$ in an external magnetic field}
\author{I.~Ya.~Korenblit}
\affiliation{School of Physics and Astronomy, Raymond and Beverly
Sackler Faculty
of Exact Sciences,\\
Tel Aviv University, Tel Aviv 69978, Israel}
\author{A.~Aharony, and O.~Entin-Wohlman}
\affiliation{Department of Physics and the Ilse Katz Center for
Meso- and Nano-Scale Science and Technology, Ben Gurion
University, Beer Sheva 84105, Israel}
\date{\today}
\begin{abstract}
The effects of quantum fluctuations on the magnetic field
dependence of the spin-wave gaps in the lamellar antiferromagnet
La$_2$CuO$_4$ are considered. Nonlinear corrections to the
spin-wave spectrum are calculated to leading order in $1/S$, where
$S$ is the localized spin.  The nearest-neighbor exchange
interactions between the Cu spins as well as the
Dzyaloshinskii-Moriya interactions are taken into account. Using
the experimental values of the components of the $g$-factor
tensor, we get a satisfactory agreement with the experimental
results for the field dependence of the gaps by Gozar {\it et al.}
[Phys. Rev. Lett. {\bf 93}, 027001 (2004)], and obtain consistent
values of the in-plane and inter-plane coupling constants. The
field dependence of the dispersion of spin waves propagating
perpendicular to the CuO$_2$ planes is also discussed.
\\[3ex]
\end{abstract}
 \maketitle

\section{Introduction}

Lanthanum cuprate, La$_2$CuO$_4$, the parent compound of the
high-temperature superconductors, is a layered weakly orthorhombic
antiferromagnet (AF), with a N\'{e}el temperature, $T_N$, of
approximately 325 K.\cite{KBS} The orthorhombic distortion is
associated with a small tilt of oxygen octahedra around each
copper ion, which also introduces the antisymmetric
Dzyaloshinskii-Moriya (DM) superexchange interaction between
neighboring spins. Because of the DM interaction, each CuO$_2$
plane acquires a small ferromagnetic moment along the $c$-axis,
perpendicular to the plane. The direction of the ferromagnetic
moments alternates in adjacent planes, and, therefore, there is no
net ferromagnetic moment in the crystal.

The DM interaction, though small, has a strong impact on the
magnetic properties of La$_2$CuO$_4$. It was established  that a
magnetic field perpendicular to the CuO$_2$ plane causes a first
order weak-ferromagnetic transition (WFT).\cite{Thio1,KBT,TA} The
critical field, $H_c$, of the transition depends on the DM
coupling as well as on the interlayer exchange. In a
non-stoichiometric sample ($T_N=234$ K), studied in Ref.
\onlinecite{KBT}, the critical field at low temperatures was
$H_c\approx 4.8$ T. In Ref. \onlinecite{Goz} the value $H_c\approx
6$ T was obtained from Raman spectrum measurements for a crystal
with $T_N=310$ K, while in a recent paper\cite{RUP}  a
significantly larger transition field of 11.5 T was found from
neutron diffraction studies for a sample with $T_N=316$ K.

Another manifestation of the DM interaction in La$_2$CuO$_4$ is
the phase diagram in an in-plane magnetic field. The
antiferomagnetic staggered moment in La$_2$CuO$_4$ is directed
along the diagonal of the CuO plaquette ($b$-axis), the in-plane
anisotropy being generated by the DM interaction. A theory
\cite{TT2} based on the mean-field version of the Hamiltonian,
which includes exchange and DM coupling between Cu spins, as well
as an out-of-plane anisotropy term, predicts that in a field along
the $b$-axis one should observe two phase transitions: a spin-flop
transition, at which the staggered moment jumps from the (b,c)
plane into the (b,a) plane ({\bf a} is perpendicular to {\bf b}
and {\bf c}) at a field $H$, determined by the DM coupling energy,
and a second transition at a higher field, when the staggered
moment rotates into the $c$-direction. Such a phase diagram was
indeed observed in Ref. \onlinecite{TT2}. Note, however, that in
Refs. \onlinecite{Goz},\onlinecite{RUP},\onlinecite{OKL} no
spin-flop transition was observed. According to these references,
the staggered moments rotate gradually from the $b$- to the
$c$-axis.

Gozar {\it et al.} \cite{Goz} used Raman spectroscopy measurements
to study the influence of a magnetic field on the in-plane
spin-wave (SW) gap in La$_2$CuO$_4$. They showed that the gap
abruptly increases at the WFT, decreases with the increase of the
field directed along the $b$- axis and increases with the field
pointing along the $a$-axis. The observed field dependence of the
gaps was described by classical theories\cite{CP,BS,BSG,LS} based
on the model, which include anisotropic symmetric and
antisymmetric intralayer coupling between the Cu spins as well as
 interlayer exchange.

 On the other hand, it is known that quantum corrections are very
 important when considering the properties of layered cuprates,
 with spin $S=1/2$.\cite{Min} Intensive calculations were performed for
  an isotropic Heisenberg two dimensional model with
   nearest neighbor interactions.\cite{Min,Ig,Be,BCP,
 ZOH,SG,Harris} Perturbative expansions in $1/S$,
 series expansions, and Monte Carlo methods were used
 to go beyond the linear SW theory.\cite{Min,Ig,Be}
 It appears that the renormalization factors obtained from the nonlinear
 SW (NLSW) theory to leading order in $1/S$ are sufficiently close to those
 which follow from more  complicated calculations even for spin
 $S=1/2$.
  The quantum effects
 renormalize  significantly the mean value of the spin at $T=0$, the
 SW stiffness, the in-plane and out-of-plane SW gaps at zero magnetic field, etc.
 In a recent paper,\cite{KK} the $1/S$ NLSW theory was employed to calculate the
  effect of quantum fluctuations on the SW
 stiffness and other parameters of La$_2$CuO$_4$,
 in a model which takes into account
  the  plaquette ring exchange. The renormalization factors which
  follow from these calculations
  and the fit of the experimental results by Coldea {\it et
  al.},\cite{CHA} appeared to be somewhat closer to unity than in the model with
  nearest-neighbor exchange only.
  In view of these results one should expect that quantum
  fluctuations are important also
  for the quantitative understanding of the SW spectrum behavior in a
  magnetic field.

  In this paper we use the NLSW theory to leading order in $1/S$ to
  calculate
  the influence of quantum fluctuations on the magnetic field
  dependence of the in-plane and out-of-plane SW gaps in
  La$_2$CuO$_4$. We show that the quantum fluctuations alter
  the expression for the critical field of the WFT,
  renormalize all physical quantities,
  which enter
  in the equations for the gaps in a magnetic field, and change therefore significantly
  the values of the spin-spin couplings extracted from the
  experiment. The theoretical results
  are in satisfactory agreement with the experimental findings.
  The theory also predicts a non-trivial effect of the magnetic
  field  on the SW propagating in the $c$-direction.

  The paper is organized as follows. In section II we present the model of magnetic
  interactions in La$_2$CuO$_4$ used in our calculations. In
  section III we derive the field dependance of the gaps for the
  field in the $c$-direction in both the linear and the nonlinear SW
  approximations. We also consider the field dependence of the dispersion of
  SW, which
  propagate in the $c$-direction. In section IV the  NLSW
  theory for the SW gaps in a magnetic field along the $b$- and
  $a$-directions is developed. In section V we compare our NLSW results
  with the experiments by Gozar {\it et al.}\cite{Goz}  Finally in
  the Appendix the Green's function are used to derive a general
  expression for the eigenvalue matrix of a model described by the
  Hamiltonian of the type (\ref{h2}). The Appendix also presents
  some useful relations between the averages of the magnon
  operators.

\section{The model}

Our Hamiltonian is written as\cite{SEA}
\begin{eqnarray}
{\cal H}&=&J\sum_{<ij>}\Bigl[{\bf S}_i\cdot{\bf S}_j- \alpha
S^z_iS^z_j\Bigr]+ \sum_{<ij>}{\bf D}_{ij}\cdot{\bf S}_i\times{\bf
S}_{j}\nonumber\\&+& J_\perp\sum_{<ik>}{\bf S}_i\cdot{\bf S}_{k}
 -\sum_i\mu_B{\bf H}\cdot\hat{\bf g}\cdot{\bf S}_i. \label{ham}
\end{eqnarray}
Here the first term describes the anisotropic exchange coupling
between nearest-neighbor spins in the CuO$_2$ layers,
 the easy plane being the crystallographic $(a,b)$
plane.  The anisotropy factor $\alpha$ in La$_2$CuO$_4$ (LCO) is
small, of order $10^{-4}$.\cite{keimer} The second  term is the DM
interaction, which leads to the canting of the spins out of the
plane. The third term is the small interlayer exchange coupling.
The sum is over the (two) nearest neighbors along the
crystallographic $c$-axis, which we choose as the $z$-axis.
Finally, the last term in Eq. (\ref{ham}) describes the  Zeeman
energy in an external field {\bf H}, $\hat{\mathbf g}$ being the
anisotropic $g$-factor; $\mu_B$ is the Bohr magneton.

The DM vector ${\bf D}_{ij}$ is perpendicular to the Cu-Cu bonds
and changes sign from one bond to the next one. In the
orthorhombic crystallographic axes ($a,b,c$), where {\bf a} and
{\bf b} point along the diagonals of the Cu plaquette, the vector
${\bf D}_{ij}$ can be written as
\begin{eqnarray}
{\bf D}_{ii+\delta_1}&=&(D,D,0)\nonumber\\
{\bf D}_{ii+\delta_2}&=&(D,-D,0), \label{dij}
\end{eqnarray}
where $\delta_1$ and $\delta_2$ are directed along the two
orthogonal bonds of the plaquette. Using these DM vectors, the DM
term in the Hamiltonian transforms into
\begin{eqnarray}
\sum_{<ij>}{\bf D}_{ij}\cdot{\bf S}_i\times{\bf S}_j&=&
\sum_{<ij>}{\bf D}\cdot{\bf S}_i\times{\bf S}_j\nonumber\\
&+& D
\sum_i\left[\sum_{\delta_1}\left(S_i^xS_{i+\delta_1}^z-S_i^zS_{i+\delta_1}^x\right)-
\sum_{\delta_2}\left(S_i^xS_{i+\delta_2}^z-S_i^zS_{i+\delta_22}^x\right)\right],
\label{DM}
\end{eqnarray}
with the vector {\bf D} of length $D$ pointing along {\bf a}.
 The axes $(x,y,z)$
are along $(a,b,c)$.

 The second term in the r.h.s. of this
expression does not contribute to the gap. It only changes
slightly, to order $D/J$, the SW dispersion.\cite{EAS} We will
therefore neglect it in what follows. Thus, for our purposes the
DM interaction can be described by the vector {\bf D}, directed
along the $a$-axis in each plaquette.

 The magnetic unit cell in the above model contains four spins,
$A,B,C,D$, in two inequivalent layers, Fig. 1. Due to the DM
interaction, these spins cant into the $c$-direction, with canting
angles, $\psi_\mu$, with $\mu=a,b,c,d$ for the spins in the unit
cell. These angles depend on the direction and on the strength of
the magnetic field.
 In zero field, they satisfy the
relation $\psi_a=\psi_b=-\psi_c=-\psi_d$, so that the
magnetic moment of the unit cell is equal to zero.
\begin{figure}
\includegraphics[width=8cm]{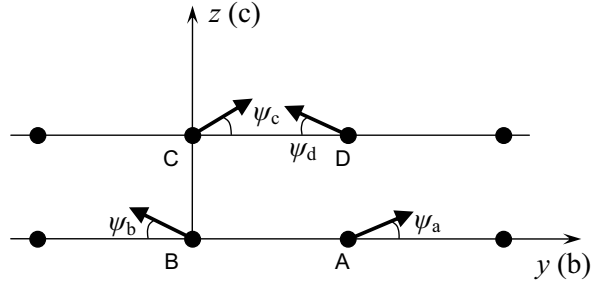}
 \caption{The four
copper spins A,B,C,D in a unit cell. Ordered spin directions are
shown by arrows. The horizontal lines refer to the CuO$_2$ layers.
The global reference axes $z$ and $y$ point along the orthorhombic
crystallographic axes $c$ and $b$ respectively.} \label{Fig. 1}
\end{figure}
To calculate the SW spectrum, it is convenient to rotate the spins
from the above global crystallographic axes $(x,y,z)$ to local
axes $(\xi,\eta,\zeta)$, with $\zeta$ along the sublattice
magnetization. When the magnetic field is in the $b$-direction,
and smaller than the spin-flop field, or in the $c$- direction,
the spins rotate in the $(y,z)$ plane.
 Then the transformation
equations from the crystallographic axes to the local ones for
spins $a$ and $b$ in the sublattices $A$ and $B$  are (see Fig. 1)
\begin{eqnarray}
S_a^y&=&S_a^\eta\sin\psi_a +S_a^\zeta\cos\psi_a,~~S_a^z=-S_a^\eta\cos\psi_a
+S_a^\zeta\sin\psi_a,\nonumber\\
S_b^y&=&S_b^\eta\sin\psi_b-S_b^\zeta\cos\psi_b,~~S_b^z=S_b^\eta\cos\psi_b
+S_b^\zeta\sin\psi_b.
\label{rot}
\end{eqnarray}
Analogous equations hold for the spins $c$ and $d$ in the second layer.

\section{Field along the \lowercase{c}-axis}

The canting angles for this direction of the field are the same
for each pair of the spins in a plane, i.e.
$\psi_a=\psi_b\equiv\psi_1$, and $\psi_c=\psi_d\equiv\psi_2$.
$\psi_1$ and $\psi_2$, and, hence, the magnetic moments of the
planes differ if $H$ is smaller than a critical field $H_c$, which
in the classical approximation is equal to \cite{Thio1}
\begin{equation}
H_c={4JJ_{\perp}S\over g_c\mu_B D}. \label{Hc}
\end{equation}
At $H=H_c$ the weak-ferromagnetic transition (WFT) happens, and
the magnetic moments of all planes became the same.\cite{Thio1} At
fields of order $H_c$, the angles $\psi_1$ and $\psi_2$ are small,
of order $10^{-2}$.

In the local framework, the Hamiltonian (\ref{ham}) has the form
\begin{equation}
{\cal H}={\cal H}_1 +{\cal H}_2 + {\cal H}_\perp. \label{hzeta}
\end{equation}
Here
\begin{eqnarray}
{\cal H}_{l=1,2}&=&-(J\cos2\psi_l+D_l\sin2\psi_l+\alpha J\sin^2\psi_l)
\sum_{<ij>}S^{\zeta}(i)S^{\zeta}(j)\nonumber\\&+&
{1\over2}\Bigl(J\cos^2\psi_l+{D_l\over2}\sin2\psi_l
-{\alpha J\over2}\Bigr)\sum_{<ij>}\Bigl[S^+(i)S^+(j) +
S^-(i)S^-(j)\Bigr]\nonumber\\
&+&{1\over2}\Bigl(J\sin^2\psi_l-{D_l\over2}\sin2\psi_l
+{\alpha J\over2}\Bigr)\sum_{<ij>}\Bigl[S^+(i)S^-(j)+
S^-(i)S^+(j)\Bigr]\nonumber\\
&-&g_c\mu_BH\sin\psi_l\Bigl[\sum_iS^{\zeta}(i)+\sum_jS^{\zeta}(j)\Bigr]
+g_c\mu_BH\cos\psi_l\Bigl[\sum_i S^{\eta}(i)-\sum_j
S^{\eta}(j)\Bigr]\nonumber\\&+&
\Bigl(J\sin2\psi_l-D_l\cos2\psi_l-{\alpha
J\over2}\sin2\psi_l\Bigr)
\sum_{<ij>}\Bigl[S^{\zeta}(i)S^{\eta}(j)-
S^{\eta}(i)S^{\zeta}(j)\Bigr],
 \label{hl}
\end{eqnarray}
where $D_l=(-1)^lD$,
\begin{equation}
S^{\pm}=S^{\xi}\pm iS^{\eta},
\label{s+-}
\end{equation}
and $i(j)$ labels the spins in the sublattice A (B).

The interlayer Hamiltonian transforms into
\begin{eqnarray}
{\cal
H}_\perp&=&-J_\perp\sum_{<ik>}\Bigl\{S^{\zeta}(i)S^{\zeta}(k)\cos\psi_{1,2}
-{1\over2}\Bigl[S^{+}(i)S^{+}(k)+S^{-}(i)S^{-}(k)\Bigr]\cos^2{\psi_{1,2}\over
2}\nonumber\\
&-&{1\over2}\Bigl[S^{+}(i)S^{-}(k)+S^{-}(i)S^{+}(k)\Bigr]\sin^2{\psi_{1,2}\over
2}+\Bigl[S^{\zeta}(i)S^{\eta}(k)-
S^{\eta}(i)S^{\zeta}(k)\Bigr]\sin\psi_{1,2}\Bigl\},
\label{hz1}
\end{eqnarray}
where $\psi_{1,2}=\psi_{1}+\psi_{2}$. We next make the
Dyson-Maleev transformation to the bosons operators $a,b,c,d$ for
the four spin operators in a unit cell. In the local framework,
this transformation is the same for all spins. It can be written
as
\begin{eqnarray}
S_\mu^+&=&\sqrt{2S}\alpha_\mu,~~S_\mu^-=\sqrt{2S}\alpha_\mu^{\dagger}\Bigl(1-
{\alpha_\mu^{\dagger}\alpha_\mu\over2S}\Bigr),\nonumber\\
S_\mu^{\zeta}&=&S-\alpha_\mu^{\dagger}\alpha_\mu,
\label{dm}
\end{eqnarray}
where the operators $\alpha_\mu$ are defined as $\alpha_1=a,~
\alpha_2=b,~ \alpha_3=c,~ \alpha_4=d$.

\subsection{Linear spin-wave theory}

It follows from Eqs (\ref{hzeta}), (\ref{hl}), (\ref{hz1}) that
the classical ground-state energy per spin, $E_0$, is
\begin{eqnarray}
E_0&=&{1\over4}[-JS^2z(\cos2\psi_1 +\cos2\psi_2)+
DS^2z(\sin2\psi_1-\sin2\psi_2)\nonumber\\&-&
2g_c\mu_BHS(\sin\psi_1+\sin\psi_2)-4J_{\perp}S^2\cos(\psi_1+\psi_2)
- \alpha JS^2z(\sin^2\psi_1 +\sin^2\psi_2)], \label{e0}
\end{eqnarray}
where $z=4$ is the number of nearest neighbors in the plane. The
minimization conditions for the energy $E_0$ with respect to
$\psi_1$ and $\psi_2$ yield
\begin{eqnarray}
 JSz(2-\alpha)\sin2\psi_1+4J_{\perp}S\sin(\psi_1+\psi_2)+
 2DSz\cos2\psi_1-2g_c\mu_BH\cos\psi_1
&=&0,\nonumber\\
JSz(2-\alpha)\sin2\psi_2+4J_{\perp}S\sin(\psi_1+\psi_2)-
2DSz\cos2\psi_2-2g_c\mu_BH\cos\psi_2 &=&0. \label{psi}
\end{eqnarray}
The solution of these equations gives $\psi_1$ and $\psi_2$ at
fields $H<H_c$. Given that $\psi_1$ and $\psi_2$ are small, and
neglecting terms of order $\alpha$ and $J_z/J$, we get
\begin{equation}
\psi_1={g_c\mu_B H-DSz\over2JSz},~~\psi_2={g_c\mu_B
H+DSz\over2JSz}. \label{psi12}
\end{equation}
With these angles, the terms in Eq. (\ref{hl}) which are proportional
to $S^\eta$, and are therefore linear in the boson operators,
cancel out.

The bilinear spin-wave Hamiltonian, which follows after the
transformation (\ref{dm}) is performed, can be written as
\begin{equation}
{\cal H}^{(2)}= \sum_{\bf q}\Bigl\{A_{\mu\nu}({\bf
q})\alpha_\mu^{\dagger}({\bf q}) \alpha_\nu({\bf q})+
{1\over2}B_{\mu\nu}({\bf q})[\alpha_\mu^{\dagger}({\bf q})
\alpha_\nu^{\dagger}(-{\bf q})+ \alpha_\mu({\bf
q})\alpha_\nu(-{\bf q})]\Bigr\}, \label{h2}
\end{equation}
where the matrices ${\bf A}({\bf q})$ and ${\bf B}({\bf q})$ are
given by
\begin{eqnarray}
A_{11}&=&A_{22}=JSz(1-2R_1)+g_c\mu_B H\psi_1 +2J_{\perp}S,
\nonumber\\
A_{33}&=&A_{44}=JSz(1-2R_2)+g_c\mu_B H\psi_2 +2J_{\perp}S,
\nonumber\\
A_{12}&=&A_{21}=JSz\gamma_{\bf q}\Bigl(R_1+{\alpha\over2}\Bigr),
\nonumber\\
A_{34}&=&A_{43}=JSz\gamma_{\bf q}\Bigl(R_2+{\alpha\over2}\Bigr),
\nonumber\\
A_{14}&=&A_{23}=A_{13}=A_{14}=0,\nonumber\\
B_{12}&=&B_{21}=JSz\gamma_{\bf q}\Bigl(1-R_1-{\alpha\over2}\Bigr),
\nonumber\\
B_{34}&=&B_{43}=JSz\gamma_{\bf q}\Bigl(1-R_2-{\alpha\over2}\Bigr),
\nonumber\\
B_{14}&=&B_{23}=2J_{\perp}Sc_z,
\nonumber\\
B_{11}&=&B_{22}=B_{33}=B_{44}=B_{13}=B_{31}=B_{24}=B_{42}=0,
\label{ab}
\end{eqnarray}
while
\begin{equation}
R_{l=1,2}=\psi_l\Bigl(\psi_l -{D_l\over J} \Bigr)=
{H^2-(DSz)^2\over4(JSz)^2},
\end{equation}
and
\begin{equation}
\gamma_{\bf q}={\cos q_x +\cos q_y\over2},~~c_z=\cos q_z.
\label{gamc}
\end{equation}
The wave-vector components in the plane and along the $z$-axis are
measured in units of the corresponding intersite distances. In Eq.
(\ref{ab}) we neglect corrections of order $J_\perp\psi_l^2$ and
$\alpha\psi_l^2$ (l=1,2).

It was shown in Ref. \onlinecite{Harris} (see also the Appendix of
this paper) that the squares of the spin-wave energy, $\omega^2$,
are the eigenvalues of the matrix
\begin{equation}
{\bf M}=[{\bf A}({\bf q})+{\bf B}({\bf q})]\times
[{\bf A}({\bf q})-{\bf B}({\bf q})].
\label{MAB}
\end{equation}
Since the magnetic field is small,
 $g_c\mu_B H\ll J$, it is not expected to influence essentially the
stiffness of the  spin-waves propagating in the $(x,y)$ plane. In
contrast, the effect of the field on the spin waves propagating in
the z-direction may be strong. We consider, therefore, only the
dispersion of the spin waves propagating in this direction, and
put $\gamma_{\bf q}=1$. The spin-wave spectrum at  $H<H_c$, which
follows then from Eqs. (\ref{ab}) and (\ref{MAB}), is given by
\begin{eqnarray}
\omega_{in}^2&=&\delta^2 +w^2\pm\sqrt{(g_c\mu_B H\delta)^2 +w^4c_z^2},\nonumber\\
\omega_{out}^2&=&\Delta^2 +(g_c\mu_B H)^2 +w^2\pm \sqrt{(g_c\mu_B
H\delta)^2 +w^4c_z^2}. \label{om<}
\end{eqnarray}
Here $\omega_{in}$ and $\omega_{out}$ are the frequencies of the
in-plane and out-of-plane spin waves, respectively. The zero-field
in-plane, $\delta$, and out-of-plane, $\Delta$, gaps
 are:
 \begin{equation}
 \delta=DSz,~~ \Delta=JSz\sqrt{2\alpha},
 \label{gap1}
 \end{equation}
 and $w^2=4zJJ_{\perp}S^2$. The lower (upper) sign in Eqs. (\ref{om<}) gives the frequencies
of the in-phase or acoustical (out-of-phase or optical)
oscillations in adjacent planes.

At $H=H_c$, the field which favors alignment of magnetic moments
in adjacent planes overcomes the interlayer coupling which favors
alternation of the moments. Then the staggered moments of all the
planes align in the same direction, accompanied by a 180$^0$
rotation
 of the spins in half of the planes.
Hence, the canting angles at $H>H_c$ are
\begin{equation}
\psi_2=\pi-\psi_1={g_c\mu_B H+DSz\over2JSz}. \label{psi>}
\end{equation}
The difference $\Delta E_0$, between the ground state energy $E_0$
at $H<H_c$ and that at $H>H_c$, is
\begin{equation}
\Delta E_0={g_c\mu_B HDS\over2J}-2J_zS^2. \label{dE}
\end{equation}
 The expression (\ref{Hc}) for $H_c$ follows from the relation
$\Delta E_0=0$.

Since the two adjacent planes are equivalent, only the two
modes which describe the in-phase oscillations in the planes
are relevant. One gets for these modes
\begin{eqnarray}
\omega_{in}^2&=&\delta^2 +g_c\mu_B H\delta +w^2(1-c_z),\nonumber\\
\omega_{out}^2&=&\Delta^2 + (g_c\mu_BH)^2 +g_c\mu_B
H\delta+w^2(1-c_z). \label{om>}
\end{eqnarray}
The spin-wave gaps which follow from Eqs. (\ref{om<}) and
(\ref{om>}) at $c_z=1$, coincide with those obtained in  different
ways in Refs. \onlinecite{CP}-\onlinecite{LS}.

\subsection{Nonlinear spin waves}

We now consider the effect of terms which are cubic and quartic in
the boson operators. It follows from Eqs. (\ref{hl}) and
(\ref{hz1}) that all the terms which are linear in $S^{\eta}$,
except those which are proportional to $H$, cancel out owing to
Eq. (\ref{psi}).  Thus, the only terms odd (cubic) in boson
operators that remain in the Hamiltonian are proportional to $H$.
For one plane and for small fields
 $g_c\mu_BH\ll JSz$ we have
\begin{equation}
{\cal H}^{(3)}=g_c\mu_BH\sqrt{1\over2S}{1\over
z}\sum_{<ij>}[a_i^\dagger a_i(b_j- b_j^\dagger)+b_j^\dagger
b_j(a_i^\dagger-a_i)]. \label{h3}
\end{equation}
The fourth-order terms of order  $1/S$ come from the exchange and
DM interactions, and from the out-of-plane anisotropy. They are
\begin{eqnarray}
{\cal H}^{(4)}_J&=&-J(1-2\psi^2_1)\sum_{<ij>}a_i^\dagger a_ib_j^\dagger b_j-
{J\over2}(1-\psi^2_1)\sum_{<ij>}a_i^\dagger b_j^\dagger(a_i^\dagger a_i+
b_j^\dagger b_j)\nonumber\\
&-&{J\over2}\psi^2_1\sum_{<ij>}a_i^\dagger a_i^\dagger a_ib_j,
\label{h4J}
\end{eqnarray}
\begin{equation}
{\cal H}^{(4)}_D={D\psi_1\over2}\sum_{<ij>}[4a_i^\dagger a_ib_j^\dagger b_j+
a_i^\dagger a_i^\dagger a_i(b_j^\dagger-
b_j)+(a_i^\dagger-a_i)b_j^\dagger b_j^\dagger b_j],
\label{hd4}
\end{equation}
\begin{equation}
{\cal H}^{(4)}_\alpha=-{J\alpha\over4}\sum_{<ij>}[( a_i-
a_i^\dagger)b_j^\dagger b_j^\dagger b_j +(b_j-
b_j^\dagger)a_i^\dagger a_i^\dagger a_i].
\label{ha4}
\end{equation}
The $1/S$ corrections to the spin wave spectrum come from the
first-order contribution of ${\cal H}^{(4)}$ and the second-order
contribution of ${\cal H}^{(3)}$. The last contribution is of
order  $H^2$, and it renormalizes the $H^2$ terms in the out-of
plane spin gap, Eqs. (\ref{om<}), (\ref{om>}). It was calculated
in Ref. \onlinecite{SM}.

To treat the quartic perturbation, we truncate all four-operator
terms by contracting out pairs of operators in all possible ways
and
 discarding the non-Hermitian terms.\cite{Harris} One finally arrives
at an effective bilinear Hamiltonian, of the form shown in Eq.
(\ref{h2}), with the matrix elements $A_{\mu\nu}$ and $B_{\mu\nu}$
depending on the following averages of the magnon operators:
\begin{eqnarray}
\nu&=& <a_i^{\dagger}a_i>=\sum_{\bf q}a_{\bf q}^{\dagger}a_{\bf q},~~
\xi=\sum_{\bf q}<a_{\bf q}b_{-{\bf q}}>\gamma_{\bf q},\nonumber\\
\eta&=&\sum_{\bf q}<a_{\bf q}^{\dagger}b_{\bf q}>\gamma_{\bf q},~~
\lambda=\sum_{\bf q}<a_{\bf q}a_{-{\bf q}}>.
\label{nueta}
\end{eqnarray}
The matrix elements $A_{\mu\nu}$ and $B_{\mu\nu}$, with $\mu=1,2; \nu=1,2$
can be written as
\begin{eqnarray}
A_{11}&=&A_{22}=JSzZ_c\Bigl[1-2\tilde{R}_1\Bigl(1+{\xi\over 2S}\Bigr)+
{\alpha\xi\over2S}\Bigr]+H\tilde{\psi_1}+2J_zSZ_m,
\nonumber\\
A_{12}&=&A_{21}=JSz\gamma_{\bf q}\Bigl[\tilde{R}_1Z_m-
{1\over 2S}(\lambda+2\eta)+Z_m{\alpha\over2}\Bigr],
\nonumber\\
B_{11}&=&B_{22}=-{Jz\over2}\Bigl(\xi \tilde{R}_1+\eta-{\alpha\xi\over2S}\Bigr),
\nonumber\\
B_{12}&=&B_{21}=JSz\gamma_{\bf q}Z_c\Bigl[1-\tilde{R}_1\Bigl(1-
{\xi\over S}\Bigr) -{\alpha\over2}\Bigl(1+{\xi\over S}\Bigr)\Bigr].
\label{AB}
\end{eqnarray}
Here $\tilde{\psi_1}$ is the tilting angle renormalized
by quantum fluctuations (see below),
\begin{equation}
\tilde{R}_1=\tilde{\psi_1}\Bigl(\tilde{\psi_1}+{D\over J}\Bigr),
\label{tR}
\end{equation}
$Z_c=1-\nu/S-\xi/S$ is the renormalization of the SW velocity,
\cite{Liu,Min} and $Z_m=1-\nu/S$ is the renormalization factor of
the average spin.

The equations for the
matrix elements $A_{\mu\nu}$ and $B_{\mu\nu}$, with $\mu=3,4; \nu=3,4$
follow from the above equations by substituting $\tilde{\psi}_2$
for $\tilde{\psi}_1$. The mixed elements $B_{14}$ and $B_{23}$ are
\begin{equation}
B_{14}=B_{23}=2J_{\perp}SZ_mc_z. \label{B14}
\end{equation}
Note that averages like $<a_ic_j>$ appear in the course of the
renormalization. They are small, since $i$ and $j$ belong to
different planes. When deriving Eqs. (\ref{AB}) we neglected also
terms of order of $\eta\psi_n$ or $\lambda\psi_n$, since (as shown
in the Appendix) $\lambda$ and $\eta$ are of order of $\psi_1^2$.

The renormalization factors $\xi$ and $\nu$
are known\cite{Min,Ig} to be equal to
\begin{equation}
\nu={1\over2}\sum_{\bf q}\Bigl({1\over\sqrt{1-\gamma_{\bf q}^2}}-1\Bigr)=
0.197,~~
\xi=-{1\over2}\sum_{\bf q}{\gamma_{\bf q}^2\over\sqrt{1-\gamma_{\bf q}^2}}
=-0.276.
\label{nuxi}
\end{equation}
This gives $Z_c=1+0.158/2S=1.158$. More accurate values of $Z_c$,
obtained from expanding to order $1/S^2$ or from Monte Carlo
simulations,
 are also available.\cite{Ig,Be}
Nevertheless, for consistency we use in what follows this value of
$Z_c$. The factors $\eta$ and $\lambda$ are given in the Appendix.
It appears that the spin-wave frequencies depend only on the sum
$\eta +\lambda$, which according to Eqs. (\ref{AB}) and
(\ref{etla}) is equal to
\begin{equation}
\lambda +\eta=-\Bigl(R_1+{\alpha\over2}\Bigr)\xi.
\label{leta}
\end{equation}

To calculate the renormalized tilting angle we add to the
classical ground-state energy (\ref{e0}) the (1/S) correction,
which follows from the bilinear Hamiltonian (\ref{h2}) after
averaging the products of the  boson operators. This yields for
the quantum corrected ground-state energy, $E$:
\begin{eqnarray}
E&=&{1\over4}\{\Bigl(1-{2\nu +\xi\over S}\Bigr)[-JS^2z(\cos2\psi_1
+\cos2\psi_2)+ DS^2z(\sin2\psi_1-\sin2\psi_2)]\nonumber\\&-&
2g_c\mu_BHS\Bigl(1-{\nu\over
S}\Bigr)(\sin\psi_1+\sin\psi_2)-4J_{\perp}S^2\Bigl(1-{2\nu\over
S}\Bigr)\cos(\psi_1+\psi_2)\}.  \label{equ}
\end{eqnarray}
We neglected here terms of order $\eta\psi_1$, as well as small
corrections which come from the out-of-plane anisotropy.

 Minimization of the energy (\ref{equ}) gives
\begin{equation}
\tilde{\psi}_{2,1}={g_c\mu_BH\over2JZ_cSz} \pm{D\over2J},
\label{psiq}
\end{equation}
in agreement with previous calculations for an AF with
$D=0$.\cite{ST,ZN} Note that the contribution to the canting angle
caused by the DM interaction is the same as in the classical
limit.

When calculating the in-plane spin-wave energy from Eqs.
(\ref{MAB}) and (\ref{AB}), one finds a contribution to the gap
which is of order $1/S$ and proportional to $H^2$. This
contribution should be cancelled out by the contribution of the
cubic terms in the Hamiltonian, since the $H^2$ term in the
in-plane gap is forbidden by symmetry. Discarding this term, we
get
\begin{eqnarray}
\omega_{in}^2&=&\tilde{\delta}^2
+\tilde{w}^2\pm\sqrt{(g_c\mu_BH)^2\Bigl(1-
{\xi\over S}\Bigr)^2\tilde{\delta}^2 +\tilde{w}^4c_z^2},\nonumber\\
\omega_{out}^2&=&\tilde{\Delta}^2 +Z_c(g_c\mu_BH)^2
+\tilde{w}^2\pm \sqrt{(g_c\mu_BH)^2\Bigl(1-{\xi\over
S}\Bigr)^2\tilde{\delta}^2 +\tilde{w}^4c_z^2}. \label{omqu<}
\end{eqnarray}
Here $\tilde{\delta}=Z_m\delta=(1-\nu/S)\delta,~
\tilde{\Delta}=Z_m\Delta, ~\tilde{w}=Z_ww$, where
$Z_w=1-\nu/S-\xi/2S\approx0.88$. As noticed above, the
renormalization of the $H^2$ term in the out-of-plane
 wave was calculated
in Ref. \onlinecite{SM}.

We see that the quantum fluctuations effectively increase the
$g$-factor in the $c$-direction. This is in contrast to the
result,\cite{BS} obtained in the limit of large $N$, $N$ being the
number of the components of the spin.

Above the WFT, when the staggered moments in all planes align in
the same direction, the angle $\psi_2$ is given by Eq.
(\ref{psiq}), while $\psi_1$ is given by $\psi_1=\pi-\psi_2$.
 It follows from Eq. (\ref{equ}) that the difference, $\Delta E$ in the
 ground state energy before and after the transition is
 \begin{equation}
 \Delta E= {g_c\mu_BHSD\over2J}\Bigl(1-{\nu\over S}\Bigr) -2J_{\perp}S^2\Bigl
 (1-{2\nu\over S}\Bigr).
 \label{dEqu}
 \end{equation}
 The condition $\Delta E=0$ then yields the transition field,
 \begin{equation}
H_c={4JZ_mJ_zS\over
g_c\mu_BD}={\tilde{w}^2\over\tilde{\delta}}\Bigl(1+ {\xi\over
S}\Bigr). \label{hcqu}
\end{equation}
The above expression for $H_c$ follows from the following simple
qualitative arguments. At the WFT the magnetic moment increases
abruptly by the value $<S_z>D/J=Z_mSD/J$ [see Eq. (\ref{psiq})],
and the gain in the magnetic energy per spin is
$<S_z>g_c\mu_BHD/J$. The loss in the exchange energy is
$2J_z<S_z>^2$. Equating these energies one arrives at Eq.
(\ref{hcqu}).

 The SW spectrum at
$H>H_c$ is
\begin{eqnarray}
\omega_{in}^2&=&\tilde{\delta}^2 +H\tilde{\delta}\Bigl(1-{\xi\over S}\Bigr)
 +\tilde{w}^2(1-c_z),\nonumber\\
\omega_{out}^2&=&\tilde{\Delta}^2 +Z_cH^2 +H\tilde{\delta}\Bigl(1-
{\xi\over S}\Bigr)+\tilde{w}^2(1-c_z).
\label{omqu>}
\end{eqnarray}

\section{Field in the $(\lowercase{a,b})$ plane}
\subsection{Field along the $b$-axis}
When a magnetic field is applied in the direction of the staggered
magnetization, i.e. along the $b$-axis, an unusual spin-flop
transition happens at the field $H=\delta$.\cite{TT2} At this
field the staggered moments rotate from the ($b,c$) plane to the
($a,c$) one, forming an angle with the $c$-axis, which decreases
with the increase of the field.
 At a higher field, equal to
$H=(\Delta^2 +2w^2 -\delta^2)/\delta$, the staggered magnetization
points in the $c$-direction.\cite{TT2}

We consider in this paper only fields smaller than $\delta$, when
the spins lie in the $(b,c)$ plane, which we defined as the
$(y,z)$ plane. As in the case of a field along the $c$-axis, the
DM interaction and the field cause the canting of the spins out of
the $(a,b)$ plane. The canting angles of the spins $a$ and $b$,
however, are no longer equal to each other.

After the rotation to the local framework is performed, we obtain
the Hamiltonian for the first plane (Fig. 1) as follows:
\begin{equation}
{\cal H}_1 = {\cal H}_{ev} +{\cal H}_{odd}.
\label{evod}
\end{equation}
Here
\begin{eqnarray}
{\cal H}_{ev}&=&-J\cos\psi_{ab}\sum_{<ij>}S^{\zeta}(i)S^{\zeta}(j)+
{J\over2}\cos^2{\psi_{ab}\over2}\sum_{<ij>}\Bigl[S^+(i)S^+(j) +
S^-(i)S^-(j)\Bigr]\nonumber\\
&+&{J\over2}\sin^2{\psi_{ab}\over2}\sum_{<ij>}\Bigl[S^+(i)S^-(j)+
S^-(i)S^+(j)\Bigr]+
g_b\mu_BH(\cos\psi_b-\cos\psi_a)\sum_iS^{\zeta}(i) \nonumber\\&+&
D\sin\psi_{ab}\sum_{<ij>}\Bigl[S^{\zeta}(i)S^{\zeta}(j)-
{1\over4}\Bigl(S^+(i)-S^-(i)\Bigr)\Bigl(S^+
(j)-S^-(j)\Bigr)\Bigr]\nonumber\\
&-&{\alpha J\over4}\sum_{<ij>}\Bigl(S^+(i)-S^-(i)\Bigr)
\Bigl(S^+(j)-S^-(j)\Bigr),
\label{hev}
\end{eqnarray}
and
\begin{eqnarray}
{\cal H}_{odd}&=&(J\sin\psi_{ab}+D\cos\psi_{ab})
\sum_{<ij>}\Bigl[S^{\zeta}(i)S^{\eta}(j)-
S^{\eta}(i)S^{\zeta}(j)\Bigr]\nonumber\\
&+&g_b\mu_BH(\sin\psi_a-\sin\psi_b)\sum_iS^{\eta}(i)
\nonumber\\&-&\alpha J
\sum_{<ij>}\Bigl[S^{\zeta}(i)S^{\eta}(j)\sin\psi_a\cos\psi_b-
S^{\eta}(i)S^{\zeta}(j)\sin\psi_b\cos\psi_a\Bigr],  \label{hodd}
\end{eqnarray}
where $\psi_{ab}=\psi_a+\psi_b$. The Hamiltonian for the second
plane can be obtained from Eqs. (\ref{hev}) and (\ref{hodd}) by
replacing $\psi_a$ and $\psi_b$ by $\psi_c$ and $\psi_d$, and
changing the sign of $D$. Finally, the interlayer coupling is
\begin{eqnarray}
H_{\perp}&=&J_{\perp}\sum_{<ik>}\Bigl\{S^\xi(i)S^\xi(k)
-\Bigl[S^{\zeta}(i)S^{\zeta}(k)+
S^{\eta}(i)S^{\eta}(k)\Bigr]\cos(\psi_a+\psi_d)\Bigr\}\nonumber\\&+&
J_{\perp}\sum_{<jk>}\Bigl\{S^\xi(j)S^\xi(k)-
\Bigl[S^{\zeta}(j)S^{\zeta}(k)+S^{\eta}(j)S^{\eta}(k)\Bigr]\cos(\psi_b+
\psi_c)\Bigr\}. \label{Jzb}
\end{eqnarray}
It follows from the above equations that the classical
ground-state energy is
\begin{eqnarray}
E_{0}&=&{1\over4}\{-JS^2z[\cos(\psi_a+\psi_b)+\cos(\psi_c+\psi_d)
\nonumber\\ &+&\alpha(\sin\psi_a\sin\psi_b
+\sin\psi_c\sin\psi_d)]-
DS^2z[\sin(\psi_a+\psi_b)-\sin(\psi_c+\psi_d)]\nonumber\\
&+&g_b\mu_BHS(\cos\psi_a-\cos\psi_b+\cos\psi_c-\cos\psi_d)\nonumber\\
&-&2J_{\perp}S^2[\cos(\psi_a+\psi_d)+\cos(\psi_b+\psi_c)]\}.
\label{eclb}
\end{eqnarray}
The minimization conditions with respect to $\psi_\mu$ yield
\begin{eqnarray}
\psi_a&=&-\psi_c={DSz(g_b\mu_BH-4J_{\perp}S-\alpha JSz)\over
\Delta^2 +2w^2-(g_b\mu_BH)^2},
\nonumber\\
\psi_b&=&-\psi_d=-{DSz(g_b\mu_BH+4J_{\perp}S+\alpha JSz)\over
\Delta^2 +2w^2-(g_b\mu_BH)^2}. \label{psiad}
\end{eqnarray}
When deriving Eqs. (\ref{psiad}) we assumed
 that $\psi_\mu$ is small,
i.e. we consider magnetic fields that satisfy the inequality
$g_b\mu_BH\ll\Delta$.

\subsubsection{Spin gaps in the one-layer model}

As we show below, the interlayer coupling
 almost does not affect the spin-wave gaps. Hence, when calculating
the gaps, one may put $J_{\perp}$ to zero,
 and consider spin waves only in one plane.
Performing the Dyson-Maleev transformation of the corresponding
terms in the Hamiltonian (\ref{hev}), one gets the bilinear and
quartic terms in the Hamiltonian as follows:
\begin{eqnarray}
{\cal H}^{(2)}&=&
\sum_{<ij>}\Bigl[JS\Bigl(1-{\tilde{\psi}_{ab}^2\over2}\Bigr)
-DS\tilde{\psi}_{ab}\Bigr](a_i^\dagger a_i+ b_j^\dagger
b_j)\nonumber\\&+&
\sum_{<ij>}\Bigl[JS\Bigl(1-{\tilde{\psi}_{ab}^2\over8}\Bigr)-{DS\over2}\tilde{\psi}_{ab}-
{\alpha JS\over2}\Bigr](a_ib_j+a_i^\dagger
b_j^\dagger)\nonumber\\&+&
{1\over2}\sum_{<ij>}\Bigl[{JS\over2}\tilde{\psi}_{ab}^2+DS\tilde{\psi}_{ab}+
\alpha JS\Bigr](a_ib_j^\dagger+a_i^\dagger b_j)+
g_b\mu_BH\sum_i(a_i^\dagger a_i-b_i^\dagger b_i), \label{h2b}
\end{eqnarray}
\begin{eqnarray}
{\cal
H}^{(4)}&=&-J\sum_{ij}\Bigl[\Bigl(1-{\tilde{\psi}_{ab}^2\over2}\Bigr)
a_i^\dagger a_ib_j^\dagger b_j
+{1\over2}\Bigl(1+{\tilde{\psi}_{ab}^2\over4}\Bigr)
a_i^\dagger b_j^\dagger(a_i^\dagger a_i+b_j^\dagger b_j )\nonumber\\
&-&{\tilde{\psi}_{ab}^2\over8}(a_ib_j^\dagger b_j^\dagger b_j+
a_i^\dagger a_i^\dagger a_ib_j)\Bigr]\nonumber\\
&+&\sum_{ij}\Bigl\{D\tilde{\psi}_{ab}a_i^\dagger a_ib_j^\dagger
b_j-{1\over4} (D\tilde{\psi}_{ab}+ \alpha
J)[(a_i-a_i^\dagger)b_j^\dagger b_j^\dagger b_j+ a_i^\dagger
a_i^\dagger a_i(b_j-b_j^\dagger)]\Bigr\}. \label{h4b}
\end{eqnarray}
Here $\tilde{\psi}_{ab}=\tilde{\psi}_{a}+ \tilde{\psi}_{b}$ is the
renormalized tilting angle. To calculate it one should add, as in
Section III, to the classical ground state energy (\ref{eclb}) the
quantum correction, i.e. the averaged SW energy, and then minimize
the full energy.

The terms linear in the boson operators in Eq. (\ref{hodd}) cancel
out when the angles given by Eqs. (\ref{psiad}) are used. The
cubic terms partly cancel out as in the case when $H$ is along the
$c$-axis. The remaining cubic terms are of second order in $H$ and
in the other small parameters of the problem, and are therefore
irrelevant. Hence, the quantum corrections can be calculated in
the Hartree approximation as was explained in the previous
Section. This gives an effective bilinear Hamiltonian, with the
coefficients $A_{\mu\nu}$ given by
\begin{eqnarray}
A_{11}(H)&=&JSzZ_c \Bigl[1-
R\Bigl(1+{\xi\over 2S}\Bigr)
 +{\alpha\over2S}\xi\Bigr]+g_b\mu_BH,\nonumber\\
A_{22}(H)&=&A_{11}(-H),\nonumber\\
A_{12}&=&{JSz\over2}\gamma_{\bf q}Z_m
\Bigl[R
-{1\over S}(\lambda+2\eta) +\alpha\Bigr],\nonumber\\
B_{11}&=&-{Jz\over4}\Bigl[\xi R
+2\eta +\alpha\xi\Bigr],\nonumber\\
B_{12}&=&JSz\gamma_{\bf q}Z_c \Bigl[1-
{R\over2}\Bigl(1-{\xi\over S}\Bigr)-
{\alpha\over2}\Bigl(1+{\xi\over S}\Bigr)\Bigr].
\label{ABquant}
\end{eqnarray}
Here
\begin{equation}
R=\tilde{\psi}_{ab} \Bigl({\tilde{\psi}_{ab}\over2} +{D\over J}\Bigr),
\label{r}
\end{equation}
and the angle $\tilde\psi_{ab}$ renormalized by quantum
fluctuations is
\begin{equation}
\tilde{\psi}_{ab}=-{2\alpha JD(Z_cSz)^2\over (Z_c\Delta)^2-
(g_b\mu_BH)^2}. \label{psiqu}
\end{equation}
Equations (\ref{psiqu}) and (\ref{r}) yield
\begin{equation}
R={\alpha(Z_c\delta)^2[Z_c^2\Delta^2-2
(g_b\mu_BH)^2]\over[Z_c^2\Delta^2- (g_b\mu_BH)^2]^2} =
{D^2\over2J^2} +O(H^4). \label{r1}
\end{equation}
Thus, the field dependence of the gaps comes only from the explicit
dependence on the field of the matrix elements
 $A_{11}$ and $A_{22}$.
We get for the in-plane, $\Omega_{in}$, and out-of-plane, $\Omega_{out}$,
gaps the following expressions
\begin{eqnarray}
\Omega_{in}^2&=&{1\over2}\Bigl[\tilde{\delta}^2+\tilde{\Delta}^2 +
2(g_b\mu_BH)^2-\sqrt{(\tilde{\Delta}^2-\tilde{\delta}^2)^2+
8(g_b\mu_BH)^2(\tilde{\delta}^2+\tilde{\Delta}^2)}\Bigr]\nonumber\\
\Omega_{out}^2&=&{1\over2}\Bigl[\tilde{\delta}^2+\tilde{\Delta}^2
+ 2(g_b\mu_BH)^2+\sqrt{(\tilde{\Delta}^2-\tilde{\delta}^2)^2+
8(g_b\mu_BH)^2(\tilde{\delta}^2+\tilde{\Delta}^2)}\Bigr]
\label{Omega}
\end{eqnarray}
The in-plane frequency $\Omega_{in}^2$ changes sign at the field
$g_b\mu_BH=\tilde{\delta}$, when the spin-flop transition takes
place.
\subsubsection{Effect of interlayer coupling}

 The generalization to a three-dimensional crystal
is straightforward. Even though $\psi_{ab}$ now depends on the
interlayer coupling $w$, the quantity $R$ is, as before, equal to
$D^2/2J^2$ up to terms of order $H^4$. The in-phase gaps therefore
do not contain $w$, and as before they are given by Eqs.
(\ref{Omega}). The spectrum for spin-waves propagating along the
$z$-axis in small fields $g_b\mu_B H\ll\Delta$ is given by
\begin{eqnarray}
\omega_{in}^2(q_z)&=&\Omega_{in}^2 +\tilde{w}^2\Bigl[1-
{4(g_b\mu_BH)^2\over\tilde{\Delta}^2-\tilde{\delta}^2}\Bigr](1\pm c_z),\nonumber\\
 \omega_{out}^2(q_z)&=&\Omega_{out}^2 +\tilde{w}^2\Bigl[1+
{4(g_b\mu_BH)^2\over\tilde{\Delta}^2-\tilde{\delta}^2}\Bigr](1\pm
c_z). \label{omega3}
\end{eqnarray}
Here the upper (lower) sign corresponds to the in-phase (out-of-phase)
modes.

\subsection{Field along the $a$-axis}

In this case the field and the DM coupling do not interfere.
Hence, the spectrum is the same as one would have for an AF in a
field perpendicular to the easy anisotropy axis.\cite{SM} The only
frequencies relevant for the acoustic spin waves in this case are
given by
\begin{eqnarray}
\omega_{in}^2(q_z)&=&\tilde{\delta}^2+Z_c(g_a\mu_B H)^2+\tilde{w}^2(1-c_z),\nonumber\\
\omega_{out}^2(q_z)&=&\tilde{\Delta}^2+\tilde{w}^2(1-c_z).
\label{oma}
\end{eqnarray}

\section{Discussion: Theory versus experiment}

Consider first the field in the $c$-direction. The value of $g_c$
in La$_2$CuO$_4$ is not well determined yet (see below). We
therefore exclude $g_c$ from the expressions for the in-plane
gaps, Eqs. (\ref{omqu<}) and  (\ref{omqu>}), using Eq.
(\ref{hcqu}), and obtain:
\begin{eqnarray}
\omega_{in}^2&=&\tilde{\delta}^2 -\tilde{w}^2\Bigl(\sqrt{H^2/H_c^2
+c_z^2}-1\Bigr),~~~H<H_c, \nonumber\\
\omega_{in}^2&=&\tilde{\delta}^2 +\tilde{w}^2(H/H_c)
+\tilde{w}^2(1-c_z),~~~H>H_c. \label{gapin}
\end{eqnarray}
These expressions differ from their classical counterparts only by
 the parameters $\tilde{\delta}$ and $\tilde{w}$, which should replace their
classical values. Thus with the proper choice of the  parameter
$\tilde{w}$ ($\tilde{\delta}$ and $H_c$ are measured in the
experiment) one can fit the experimental data by Gozar {\it et
al.}\cite{Goz} in the same way as was done in Refs.
\onlinecite{Goz}, \onlinecite{BSG} and \onlinecite{LS}. According
to Ref. \onlinecite{Goz}, the derivative
$d\omega^2_{in}(H>H_c)/dH$ is given by
\begin{equation}
{d\omega^2_{in}(H>H_c)\over dH}={\tilde{w}^2\over H_c}=0.35
{(meV)^2\over T}. \label{dom}
\end{equation}
There is an uncertainty in the value of $H_c$, because of the
hysteresis observed in the first-order WFT. We choose the value
$H_c=6.3$ T, at which the gap ceases to decrease with the increase
of the field.\cite{Goz} With this value of $H_c$ we obtain from
Eq. (\ref{dom}) that $\tilde{w}^2=4Z_w^2JJ_{\perp}=2.19$
(meV)$^2$. Given that $J=135$ meV,\cite{keimer} we get:
\begin{equation}
J_{\perp}=5.3\cdot10^{-3} {\rm meV}, ~~~
\alpha_{\perp}=J_{\perp}/J= 3.9\cdot10^{-5}.\label{Jperp}
\end{equation}
The ratio $\tilde{w}^2/H_c$ can also be obtained from the relation
\begin{equation}
\tilde{w}^2/H_c=\tilde{\delta}g_c\mu_B(1-\xi/S),\label{wHc}
\end{equation}
 if the $g$-tensor is known.

According to the calculations performed in  Ref. \onlinecite{KHH},
the principal values of the $g$-tensor are: $g_c=2.45,~
g_a=g_b=2.11$. Similar values were later obtained in Ref.
\onlinecite{LS}. The experiment gave somewhat lower values:
$g_c=2.3,~ g_a=g_b=2.08$,\cite{KHH} or $g_c=2.24,~
g_a=g_b=2.06$.\cite{EKS} In what follows we use the last values of
the $g$-tensor. Given that $\tilde{\delta}$ is equal to 2.16
meV,\cite{Goz} we have: $\tilde{w}^2/H_c=0.43$ (meV)$^2$/T. This
value is somewhat larger than that, which follows from the fit of
the field dependence of the gap, Eq. (\ref{dom}). The discrepancy
might be caused by the insufficient accuracy of the $1/S$
expansion, since the factor $1-\xi/S=1.55$ is significantly larger
than one.

With  $\tilde{\delta}= 2.16$ meV (2.3 meV according to Ref.
\onlinecite{keimer}) we obtain the DM coupling as:
$D=\tilde{\delta}/2Z_m=1.78$ meV (1.90 meV). This value is by a
factor two larger than that obtained in Ref. \onlinecite{Goz}.
Note that the renormalization of both the in-plane and
out-of-plane gaps is given by $Z_m$ rather than $Z_c$ used in
Refs. \onlinecite{Goz} and \onlinecite{keimer}. For more
discussion on this subject see Ref. \onlinecite{Kim}.

The dispersion
of SW propagating in the $c$-direction also depends strongly on
the field. We have at $H<H_c$
\begin{equation}
\omega_{in}^2(q_z)-\omega_{in}^2(0)=\tilde{w}^2\Bigl(\sqrt{1+H^2/
H_c^2}- \sqrt{c_z^2+H^2/ H_c^2}\Bigr). \label{omqz}
\end{equation}
The dispersion decreases with the increase of the field, and at
the critical field, $H=H_c$, it jumps to the value in zero field,
\begin{equation}
\omega_{in}^2(q_z)-\omega_{in}^2(0)=\tilde{w}^2(1-c_z).\label{omqz1}
\end{equation}
At fields larger than $H_c$ the dispersion does not depend on the
field.

When the field is in the $b$-direction, the measured in-plane gap
decreases with the increase of the field as\cite{Goz}
$\omega_{in}^2=\tilde{\delta}^2-\gamma_b H^2$, with
$\gamma_b=0.025$ (meV/T)$^2$. This dependence follows from Eq.
(\ref{Omega}) for fields small in comparison with the out-of-plane
gap $\Delta$. In this case we have
\begin{equation}
\gamma_b={1+3(\tilde{\delta}/\tilde{\Delta})^2\over1-
(\tilde{\delta}/\tilde{\Delta})^2}(g_b\mu_B)^2. \label{gama}
\end{equation}
Given $g_b=2.06$, $\tilde{\delta}=2.16$ meV, and the above value
of $\gamma_b$, we get $\tilde{\Delta}=5.3$ meV, in good agreement
with the neutron measurement result $\tilde{\Delta}=5.5$
meV.\cite{keimer} It follows from this result that the
out-of-plane anisotropy parameter $\alpha$ is
\begin{equation}
\alpha=\tilde{\Delta}^2/8J^2Z_m^2=5.2\cdot10^{-4}.\label{alpha}
\end{equation}
The value of $\alpha$ is by an order of magnitude larger than
$\alpha_{\perp}$. This implies that $T_N$ in lanthanum cuprate is
determined mainly by the out-of-plane anisotropy.

Finally, we consider the gap when the magnetic field points in the
$a$-direction. The experimental findings were fitted in Ref.
\onlinecite{Goz} by the relation
$\omega_{in}^2=\tilde{\delta}^2+\gamma_a H^2$, with
$\gamma_a=0.015$ (meV/T)$^2$. The theoretical value, which follows
from Eq. (\ref{oma}), with $g_a=2.06$, is somewhat larger:
$\gamma_a=0.0165$ (meV/T)$^2$. Note that in the model, which
includes ring exchange, $Z_c$ was found to be $Z_c=0.96$.\cite{KK}
With this value of $Z_c$ one gets $\gamma_a=0.014$ (meV/T)$^2$.

In conclusion, we calculated the effect of quantum fluctuations on
the SW spectrum in La$_2$CuO$_4$ in an external magnetic field.
With the experimental values of the g-tensor, the in-plane gap,
and the WFT critical field,  a satisfactory agreement of the
theory with the experimental findings was obtained. Given the
renormalized values of the parameters, which determine the field
dependence  of the SW gaps, we got new values of the in-plane DM
and inter-plane couplings, and of the in-plane anisotropy
$\alpha$.

\begin{acknowledgements}
 We acknowledge helpful discussions with A.~B.~Harris
 and G.~Blumberg. We would like to thank A. Gozar for sending us
 a preprint of his paper prior to publication.
This work was supported by a grant from the German-Israeli
Foundation (GIF).
\end{acknowledgements}
\appendix*
\section{}
Define the Green's functions
\begin{equation}
G_{\mu\nu}({\bf q},t)=
<<\alpha_\mu({\bf q},t),\alpha_\nu^{\dagger}({\bf q})>>,~~
F_{\mu\nu}({\bf q},t)=
<<\alpha_\mu^{\dagger}({\bf q},t)\alpha_\nu^{\dagger}(-{\bf q})>>.
\label{gf}
\end{equation}
These functions obey the following equations of motion:
\begin{eqnarray}
i{d\over dt}G_{\mu\nu}({\bf q},t)&=&\delta(t)\delta_{\mu\nu}+
<<[\alpha_{\mu},{\cal H}],\alpha_\nu^{\dagger}({\bf q})>>\nonumber\\
i{d\over dt}F_{\mu\nu}({\bf q},t)&=&<<[\alpha_{\mu}^{\dagger}(-{\bf q}),
{\cal H}],\alpha_{\nu}^{\dagger}({\bf q})>>.
\label{gfeq}
\end{eqnarray}
Given the Hamiltonian (\ref{h2}), one gets for the  Fourier
transform of the functions {\bf G} and {\bf F}
\begin{eqnarray}
\Bigl(\omega{\bf I}-{\bf A}({\bf q})\Bigr){\bf G}(\omega,{\bf q})
-{\bf B}({\bf q})
{\bf F}(\omega,{\bf q})&=&{\bf I},\nonumber\\
\Bigl(\omega{\bf I}+{\bf A}({\bf q})\Bigr){\bf F}(\omega,{\bf q})
+{\bf B}({\bf q}){\bf G}(\omega,{\bf q})&=&0,
\label{GF}
\end{eqnarray}
where ${\bf I}$ is a unit matrix.

 The equations of the eigenvalue
problem can be written as
\begin{eqnarray}
(\omega{\bf I}-{\bf A})|G>-{\bf B}|F>&=&0,\nonumber\\
(\omega{\bf I}+{\bf A})|F>+{\bf B}|G>&=&0,
\label{eigen}
\end{eqnarray}
where $|G>$ and $|F>$ are  columns of the corresponding matrices.

This gives
\begin{eqnarray}
\omega|G-F>&=&({\bf A}+{\bf B})|G+F>,\nonumber\\
\omega|G+F>&=&({\bf A}-{\bf B})|G-F>.\nonumber\\
\label{eigen1}
\end{eqnarray}
Hence,
\begin{equation}
\omega^2{\bf I} -({\bf A}+{\bf B})({\bf A}-{\bf B})=0.
\label{eigen2}
\end{equation}
Thus, the squares of the
spin-wave energy, $\omega^2$, are the eigenvalues of
the matrix \cite{Harris}
\begin{equation}
{\bf M}=[{\bf A}({\bf q})+{\bf B}({\bf q})]\times
[{\bf A}({\bf q})-{\bf B}({\bf q})].
\label{MAB1}
\end{equation}
The quantities $\eta$ and $\lambda$, defined in Eq. (\ref{nueta}),
are related to the Green's functions as
\begin{eqnarray}
\eta&=&{1\over\pi}\int d\omega \Im\sum_{\bf q}G_{21}({\omega,\bf
q}) \gamma_{\bf q},
\nonumber\\
\lambda&=&{1\over\pi}\int d\omega \Im\sum_{\bf
q}F_{11}({\omega,\bf q}). \label{etal}
\end{eqnarray}
Together with Eqs. (\ref{GF}), this leads to the relations
\begin{eqnarray}
\eta&=&-{A_{12}\over2JSz}\sum_{\bf q}{\gamma_{\bf q}^4\over(1-
\gamma_{\bf q}^2)^{3/2}},\nonumber\\
\lambda&=&{A_{12}\over2JSz}\sum_{\bf q}{\gamma_{\bf q}^2\over(1-
\gamma_{\bf q}^2)^{3/2}}.
\label{etal1}
\end{eqnarray}
Thus, we have
\begin{equation}
\eta+\lambda={A_{12}\over JSz}\xi. \label{etla}
\end{equation}

\end{document}